\documentclass{article}
\usepackage{pifont}
\usepackage{amsmath}
\usepackage{paralist}
\usepackage{graphicx}
\usepackage{float}
\usepackage{appendix}

\usepackage{graphicx}
\usepackage{subfigure} 
\usepackage{array}
\usepackage{booktabs} 
\usepackage{multirow}
\usepackage{setspace}

\usepackage{caption}
\DeclareCaptionLabelFormat{cont}{#1~#2\alph{ContinuedFloat}}
\newcommand{\tabincell}[2]{\begin{tabular}{@{}#1@{}}#2\end{tabular}}

\usepackage[left=1.3in,top=0.6in,right=1.3in,bottom=0.8in]{geometry} 

\title { \textbf{ Influence of Murder Incident of Ride-hailing Drivers  on   Ride-hailing User's  Consuming Willingness in Nanchang}}
\author{Guangxin He, Shenghuan Yang, Miaomiao Lei, Xing Wu, Yixin Sun, Yimeng Dang \\ 
Jiangxi University of Finance and Economics\\
\small{This paper is awarded \textbf{First Prize} of}\\ 
\small{“Huachuang Cup” Market Survey and Analysis Competition in China}}
\date{}
\begin{document}
\maketitle  
 \rmfamily
\begin{abstract}
Due to the frequent murder incidents of ride-hailing drivers in China in 2018, ride-hailing companies took a series of measures to prevent such incidents and ensure ride-hailing passengers' safety. This study investigated users' willingness to use ride-hailing apps after murder incidents and users' attitudes toward Safety Rectification. We found that murder incidents of ride-hailing drivers had a significant adverse impact on people's usage of ride-hailing apps. Female users' consuming willingness was 0.633 times that of male users, such as" psychological harm" was more evident among females, and Safety Rectification had a calming effect for some users. Finally, we found that people were satisfied with ride-hailing apps' efficiency, but were not satisfied with safety and reliability, considered them important; female users were more concerned about the security than male users.

\end{abstract}
\newpage
\section{Introduction}

May 6, 2018, a 21-year-old air stewardess was killed by her driver when she used Didi Chuxing's ride-hailing service in Zhengzhou, capital of China's Henan Province. Three months after the murder, another ride-sharing passenger was raped and killed by a driver of Didi Hitch, a car-pooling service offered by China’s largest ride-hailing company Didi Chuxing. Since October 2015, drivers hired by the ride app have committed a total of 39 crimes, ranging from minor traffic violations to more severe cases of intentional injury, drug trafficking and rape. With ride-hailing become an increasingly important part of people's lives as an affordable alternative means of getting around, ride-hailing companies should take measures to ensure people's safety and remain the development of online ride-hailing service. Since the death of the young girl on August 24, Didi Chuxing's ride-hailing service has been condemned by people from all walks of life. Then Didi suspended its carpooling service from 10 p.m. to 6 a.m. and began to use video and audio recording during ride trips to warn drivers not to commit crimes, what is more, they also improved service quality and optimized ride-hailing apps. 

At present,murder incidents of ride-hailing drivers have drawn many people's attention, the main purpose of this market analysis competition is to do further research for current affairs. Therefore, in this study, we investigate people’s attitude of consuming willingness for ride-hailing after murder of ride-hailing drivers and Safety Rectification.

Ride-hailing has attracted attention from all walks of life. Since 2011, the private ride-hailing companies Uber and Lyft have expanded into more and more US cities. The impacts of ride-hailing services on the transportation system have been immediate and major. Sadowsky, Nicole and Nelson, Erik (2017), The joint presence of the two major private ride-hailing services transformed ride-hailing services from a public transportation complement to a public transportation substitute. Also, Young M and Farber S (2018), the rise of ride-hailing corresponds to a significant decrease in taxi ridership and a rise in active modes of travel. This proves that car-hailing services have gradually become an important mode of transportation in people’s daily lives. Laize Andréade Souza Silva(2018), the results show that the majority of ridesourcing trips is replacing taxi and public transport trips. What’ more, Safety and cost are the main reasons that influence the decision of sharing trips via ride-splitting.

Ride-hailing services have rapidly gained in popularity and grown extensively, some problems also arise. Chenglong C and Xiaoling Z (2018), since the service is offered over the Internet, there is a great deal of uncertainty about security and privacy. Raise significant privacy concerns, because service providers are able to track the precise mobility patterns of all riders and drivers. Not only that, some social problems also let us realize that users' privacy and security also have a series of problems, which need to be actively solved by the platform. Ma L and Zhang X(2019), users’ perception of physical risk is the most important factor negatively affecting users’ trust in drivers.Also, users’ trust in drivers has a positive effect on users’ trust in the ride service platform and their attitude towards the platform. Therefore, we should come up with new schemes to enhance privacy protection.

Pham A, Dacosta I, Jacot-Guillarmod B, et al(2017), it propose PrivateRide, a privacy-enhancing and practical solution that offers anonymity and location privacy for riders, and protects drivers’ information from harvesting attacks. Chenglong C and Xiaoling Z (2018), focusing on the uncertainty about security and privacy, we changed payment pattern of existing systems and designed a privacy protection ride-hailing scheme. E-cash was generated by a new partially blind signature protocol that achieves e-cash unforgeability and passenger privacy. Particularly, in the face of a service platform and a payment platform, a passenger is still anonymous. Additionally, a lightweight hash chain was constructed to keep e-cash divisible and reusable, which increases practicability of transaction systems.
\section{Data}
Utill 2018, there is no available research data reflecting the public's attitude towards murder incidents of ride-hailing drivers in China, so we need to collect  data by ourselves. In this study, we select people in Nanchang as the object of this survey. In order to meet the representativeness and randomness of sample, we carried out stratified sampling of each district in Nanchang. Taking into account factors such as travel costs, time, energy, and survey topics,we select people in Nanchang as the object of this survey, and  issue self-filled questionnaires to  obtain data. 

\subsection{Questionnaire Design }

In the process of designing the questionnaire, we read relevant domestic and foreign literature to build a questionnaire framework . At the same time, we followed the principle of highlighting the key points while taking into consideration the overall situation of the  survey area and survey participants, and determined three perspectives of survey following as : respondent’s basic information(gender, occupation, age); attitude towards the murder of ride-hailing drivers; attitude towards Safety Rectification; attitude towards the importance and satisfaction of ride-hailing software.
\subsubsection{Pre-survey}
In order to improve the setting of questionnaires, check for defects, and determine the sample size, we went to places with high traffic volume, such as Bayi Square and Qiushui Square, and pre-surveyed 50 questionnaires for passersby.
Through the analysis of recycling questionnaires, we found that the questionnaire did not clearly define the boundaries of age and occupation, and lacked some connection questions. Through the communication with our advisor and several revisions, we finally determined the final questionnaire.

In our survey, although sampling errors are inevitable, they can be measured and controlled by changing the sample size. Under the same conditions, the larger the sample size, the smaller the sampling error. According to the sample size calculation formula, the sample size of the required sample is: 
\begin{equation*}
    n=\frac {{\left( Z_{\alpha/2}\right) }^{2} {\sigma}^{2}  }{E^{2}}
\end{equation*}

With the confidence level of $1-{ \alpha}$ at 95\%, the estimated error $ E = 0.05$. At the beginning of implementation, our team took a pre-survey to  check if there is any unreasonable point in the questionnaire in order to conduct the investigation smoothly in the formal survey. Among the 50 questionnaires issued, 48 were valid. From table\ref{1}, We take the mean value of question 19 (attitude towards the importance of various performances of ride-hailing software) in the questionnaire as the sample mean, and we get the variance of the sample drawn as      ${\overset{\frown}{\sigma}}^{2}=0.73$.
According to the formula, the adjusted sample size was determined to be 1121. Considering the inevitable non-response phenomenon in the sample survey, the final sample size was determined to be 1200 in the formal survey.
\begin{table}[H]
\tiny
\centering

\centering
\caption{Question 19 Analysis}
\label{1}
\resizebox{90mm}{18mm}{

\begin{tabular}{ccccccc}

		\hline\hline\noalign{\smallskip}
		
		 & N & Min& Max& Mean&  \tabincell{c}{ Standard \\deviation }&variance \\
	\noalign{\smallskip}\hline\noalign{\smallskip}
	App quality& 48 & 2& 5 &3.63 & 0.866 & 0.75\\
		\specialrule{0em}{1.5pt}{1.5pt}
			Car Setting &48 & 1 & 5 & 3.5 &0.744 & 0.553  \\
			\specialrule{0em}{1.5pt}{1.5pt}

	\tabincell{c}{ Driver service\\ attitude }	 &48 & 2& 5&4.27 &0.869 & 0.755  \\

       Ride Expenses &48 & 1 & 5 & 3.92 &0.821 & 0.674
       
       \\	
       
       Passenger safety &48 & 2&5 & 4.33&0.93 & 0.865 \\
      \tabincell{c}{ Passenger Information\\ Security }  & 48 &1 &5 &4.35 &0.863 & 9.733 \\
       \tabincell{c}{ Driver registration\\Criteria }  &48 & 1&5 & 4.33 &0.883 &0.78  \\
         Driver  Identity Check &48 &2 &5 & 4.4 &0.869 & 0.755 \\

		\noalign{\smallskip}\hline\hline
	
	\end{tabular}
	}
	\end{table}

In this survey, we used  stratified sampling method to issue  questionnaires in each district according to  the population proportion of each district given in  Nanchang Statistical Yearbook of 2017.

\begin{table}[H]
\centering

\centering
\caption{Questionnaire Distribution}
\label{222}
\resizebox{\textwidth}{18mm}{

\begin{tabular}{ccccccccc}

		\hline\hline\noalign{\smallskip}
		Area & \tabincell{c}{East Lake \\District} &\tabincell{c}{West Lake \\District}&\tabincell{c}{ Qingyunpu\\District }&\tabincell{c}{ Qingshan Lake\\District }&\tabincell{c}{ Red Valley Beach \\District }&\tabincell{c}{ Gaoxin \\District }&\tabincell{c}{ Jingkai \\District }&\tabincell{c}{Total} \\
		\noalign{\smallskip}\hline\noalign{\smallskip}
		\tabincell{c}{Population \\ (in thousands) total } & 470 & 450& 270 & 430 & 230 & 250 & 140 &2240\\
		\specialrule{0em}{1.5pt}{1.5pt}
			\tabincell{c}{Sample Proportion} &20.98\% & 20.89\% & 12.05\% & 19.19\% &10.26\% & 11.16\% & 6.25\% & 100\%  \\
			\specialrule{0em}{1.5pt}{1.5pt}

			\tabincell{c}{Sample Size} &251 & 250& 143 &228 &121 & 132& 75 & 1200  \\
		\specialrule{0em}{1.5pt}{1.5pt}	
			
       \tabincell{c}{Valid Questionnaires } &245 & 242 & 139 & 222 & 119 & 130 & 70 &1167\\
		\specialrule{0em}{1.5pt}{1.5pt}	
			
       \tabincell{c}{Effective returns-ratio  } &97.6\% & 96.8\% & 97.2\% & 97.36\% & 98.34\% & 98.48\% & 93.3\% & 97.25\%
       
       \\

		\noalign{\smallskip}\hline\hline
	
	\end{tabular}
	}
	\end{table}

\subsubsection{	Questionnaire Process}
At the beginning, we took the form of online and offline questionnaire surveys. Due to the uncertainty of online questionnaires, we decided to use all offline questionnaires. When processing the questionnaire, we follow these steps:
\begin{itemize}
\item  Identify the validity of questionnaires, 
\begin{compactitem}
\setlength{\itemsep}{4pt}
\item[\ding{226}] The questionnaire which is completed is  valid questionnaire.
\item[\ding{226}]  The questionnaire which has a lot of information  is considered invalid;
\item[\ding{226}] A few questions missed, the questionnaire  considered as remedial questionnaire.
\end{compactitem}
\item  Remedial questionnaire processing
\begin{compactitem}
\setlength{\itemsep}{4pt}
\item[\ding{226}] The missing values in each variable were replaced by the mean of that variable. In our study, this method of data imputation performed better then substituting the missing values with 0.
\end{compactitem}
\item   Input data into analysis software\\
To input data into SPSS software, we follow these principles:
\begin{compactitem}
\setlength{\itemsep}{4pt}
\item[\ding{226}] Visualize the data.
\item[\ding{226}] Guarantee data originality and integrity.
\item[\ding{226}] Ensure no missing value.
\end{compactitem}
\item  Validity and reliability of questionnaires analysis

After finishing pre-survey, we exploit Cronbach's $\alpha$ to test the validity and reliability of the questionnaire. Cronbach's $\alpha$ reliability coefficient is  the most commonly used to test  validity and reliability of the questionnaire, and it is generally believed that the reliability of questionnaires should be higher than 0.7.We obtain Alpha reliability coefficient $\alpha$ = 0.84, therefore the questionnaire is valid and reliable.
\end{itemize}

\subsubsection{Questionnaire System}
Through pre-survey, we modified some unreasonable questions in the questionnaire and determined the final questionnaire. We finally set up three aspects of the questionnaire: Passenger's basic information; Incident and service improvement impact analysis;  Passenger Evaluation Analysis. Passenger Evaluation Analysis includes basic information(gender, age, occupation) and usage of ride-hailing(common ways to go outside, used ride-hailing brand , usage frequency).  Incident and service improvement impact analysis includes incident impact Analysis (attention, attitude to the cause of the incident, impact on passengers’ usage of ride-hailing) and safety rectification impact analysis(understanding degree, views on rectification, impact of rectification on passenger, changes in passenger satisfaction before and after rectification). Passenger Evaluation Analysis includes degree of satisfaction and Importance of service quality.
\begin{figure}[H]
	\centering
     
     \label{ Questionnaire System }
		\includegraphics[width=4in]{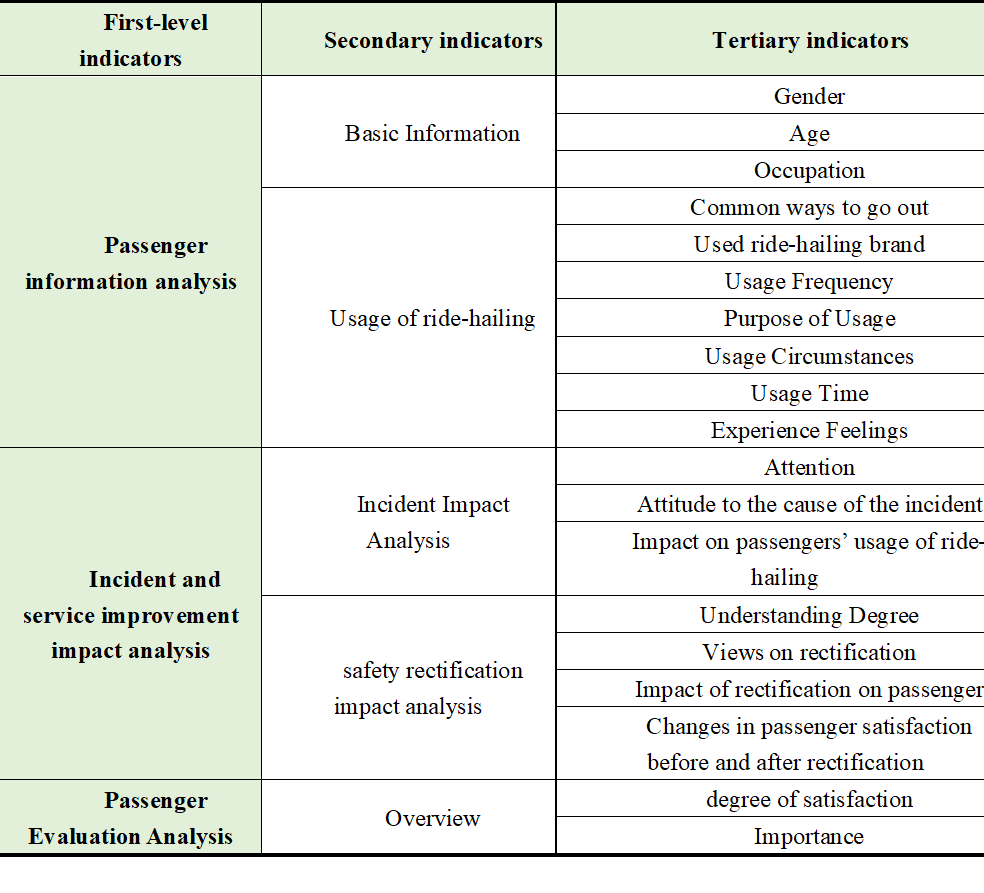}	
			\caption{Questionnaire System}
		\label{ Questionnaire System }

	\end{figure}

\section{	Passenger Basic Information Analysis}

 \begin{figure}[H]
	\centering

	\subfigure[Age]{
		\begin{minipage}{0.5\linewidth}
			\centering
			\includegraphics[width=2.2in]{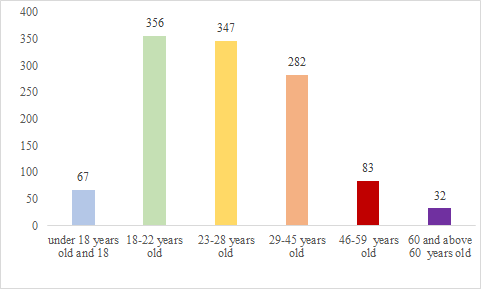}
		\end{minipage}%
	}%
	\subfigure[Occupation]{
		\begin{minipage}{0.5\linewidth}
			\centering
			\includegraphics[width=2.2in]{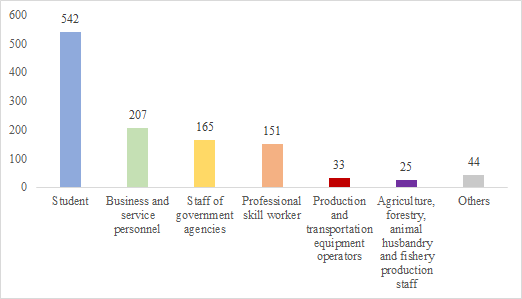}
		\end{minipage}%
	}%
	\\
	\subfigure[Gender]{
		\begin{minipage}{0.5\linewidth}
			\centering
			\includegraphics[width=2.2in]{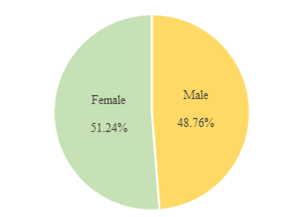}
		\end{minipage}%
	}%
		\caption{Personal Characteristics}
			\label{23}

\end{figure} 
Figure \ref{23}(a) shows that the respondents are mainly in the age group of 18-45, occupies 84.4\% of the total.  According to (b), we can find the objects of investigation are mainly students and persons engaged in business and service industries. Due to people in other occupations are not usually easily accessible, this investigation is consistent with the facts. As for (c), it shows that the male to female ratio is more balanced, consistent with the population distribution of the male to female ratio in Nanchang.

\section{Analysis of ride-hailing use}
\subsection{Transportation Choice}	
\subsubsection{	Overall Transportation Choice}
\begin{figure}[H]
	\centering

		\includegraphics[width=4in]{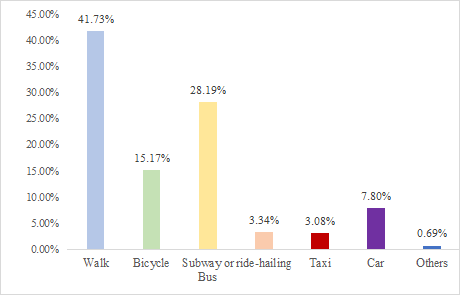} 
		\caption{ Transportation }	
	 \label{ Transportation }

	\end{figure}

Figure \ref{ Transportation } shows that 41.73\% of people’s main mode of transportation is walking, and most people are willing to walk not too far away. The main mode of transportation for 28.19\% of people is bus and subway, which are still indispensable means of transportation. So despite the high popularity of online taxis, they are still not the first choice for people to travel.

\subsubsection{Differences in Transportation Choices by Gender}
\begin{figure}[H]
	\centering

		\includegraphics[width=4in]{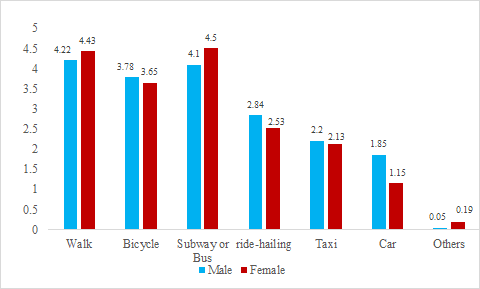}	
		\caption{ Transportation and Gender } 
		\label{nn}
	\end{figure}

Figure \ref{nn} shows that women are more inclined to choose public transportation,  such as walking, bus or subway, and pay attention to travel safety issues. More men choose to ride, taxi, private car and bicycle than women, and pay more attention to the comfort of riding. From the figure, we can see the differences between men and women in the choice of transportation.
\subsubsection{Differences in Transportation Choices in Occupation
}
\begin{figure}[H]
	\centering

		\includegraphics[width=4in]{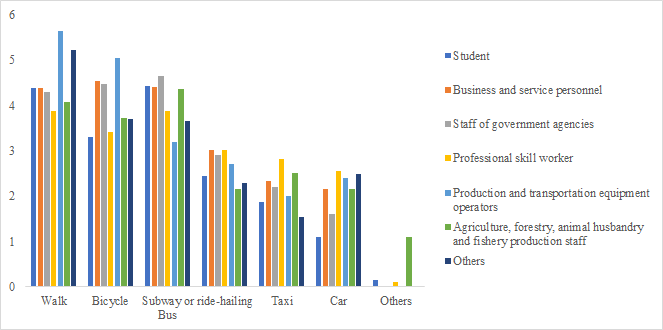}
		\caption{ Transportation and Occupation }
		\label{55}
	\end{figure}
	
Figure \ref{55}  presents that students are more inclined to choose public transport and pay more attention to travel safety. Commercial, service industry personnel, government staff, and other office workers tend to travel more conveniently, quickly, and economically. The agricultural, forestry, animal husbandry and sideline fishery production personnel are mainly inclined to walk and cycle. Overall, the main means of transportation for people to travel is still subways and buses, which shows that convenience and economy are important factors for people to consider when traveling.

\newpage
\subsection{Usage of Ride-hailing Brands}

\begin{figure}[H]
	\centering
 
		\includegraphics[width=4in]{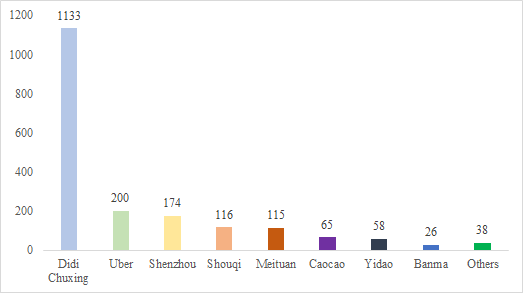}	
		\caption{ Ride-hailing Brands }
		 \label{99}

	\end{figure}
Figure \ref{99} presents that Didi Chuxing has occupied more than 90\% of the market share among the respondents. Then, 17.14\% of the respondents have used Uber, and 14.91\% have used the Shenzhou car. As a rising star, Didi Chuxing has became the largest ride-hailing company in China.
\subsection{Ride-hailing Frequency Analysis}

\begin{figure}[H]
	\centering
   
     \label{oo}
		\includegraphics[width=2.5in]{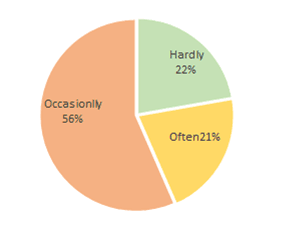}
		\caption{ Use Frequency }
		 \label{oo}
		
	\end{figure}
Figure \ref{oo} shows that most people only use taxi services occasionally, that is, only use in certain situations. Ride-hailing has not become a daily mean of transportation for people. The ride-hailing service can continue to expand potential market.However, ride-hailing company in China are facing  a severe test. A series of murder incidents have made people more afraid to ride by using ride-hailing service. Therefore, the company must take measures to ensure the safety of passengers to gain people's trust.

\subsection{Ride-hailing Service Use}
 \begin{figure}[H]
	\centering

	\subfigure[Use Reason]{
		\begin{minipage}{0.5\linewidth}
			\centering
			\includegraphics[width=2.2in]{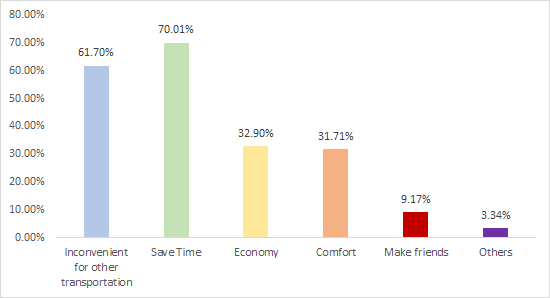}
		\end{minipage}%
	}%
	\subfigure[Use Range]{
		\begin{minipage}{0.5\linewidth}
			\centering
			\includegraphics[width=2.2in]{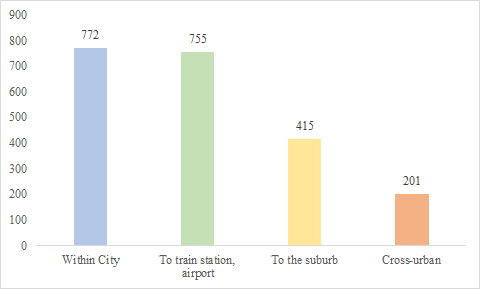}
		\end{minipage}%
	}%
	\\
	\subfigure[Use Time]{
		\begin{minipage}{0.5\linewidth}
			\centering
			\includegraphics[width=2.2in]{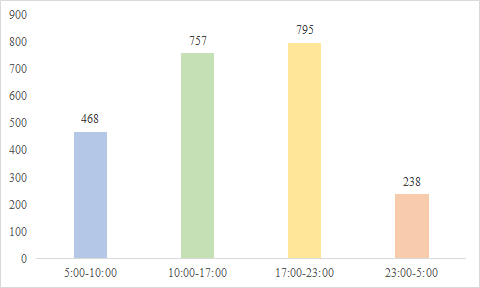}
		\end{minipage}%
	}%
		\caption{Ride-hailing Use}
			\label{ ride-hailing Use }

\end{figure}

Figure \ref{ ride-hailing Use }(a) shows that 70\% of people take a ride for saving time, and people who are inconvenient to use other means of transportation to use online car-hailing account for 61.70\%, which show some of the advantages of online car-hailing are time-saving, convenience and car use anytime and anywhere. And for (b) and (c), people use ride-hailing more frequently in urban areas and they usually take a ride during the day.

\subsection{Ride-hailing experience}

\begin{figure}[H]
	\centering
     
     \label{ Ride experience }
		\includegraphics[width=4in]{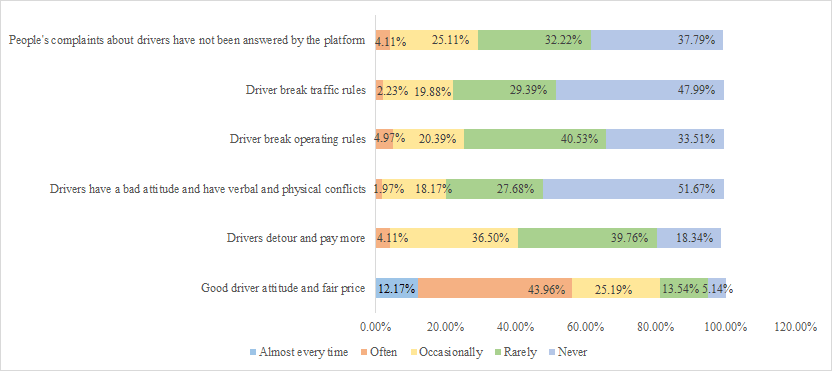}
		\caption{ Ride Experience }	
		
		 \label{ Ride experience }
	\end{figure}
Figure \ref{ Ride experience } shows that most drivers are friendly and enthusiastic, but there are also cases where drivers deliberately drive around the road and charge more, which conflicts with passengers. At the same time, ride-hailing companies should strengthen supervision and actively feedback passengers' complaints.
\newpage
\section{Murder Incident of ride-hailing drivers Impact Analysis}
\subsection{ Chi-square Test of Personal Characteristics and Incidents Attention } 
 Chi-Square Test of Independence is used to determine if there is any significant relationship between two variables. Therefore we exploit Chi-Square Test to find whether age, gender and occupation have significant relationship with people's attention to murder incidents.To analyze the difference in  attention to murder incidents of ride-hailing drivers under different ages, genders, and occupations, we also used Cross-Sectional Analysis to analyze the attention degree of different age,gender and occupation. Table \ref{chi} illustrates that the P values of age and occupation are  less than 0.05 at a significant level of 5\%, indicating that people under  different age, occupation  pay different the degree of attention to murder incidents of ride-hailing drivers . However, male and female have no difference, because those extreme incidents have drawn many people’s attention regardless of gender.\\
\begin{table}[H]
	\centering
     
     \caption{Attention to Murder Incidents Chi-square Test }
		
		  \label{chi}
		\includegraphics[width=4in]{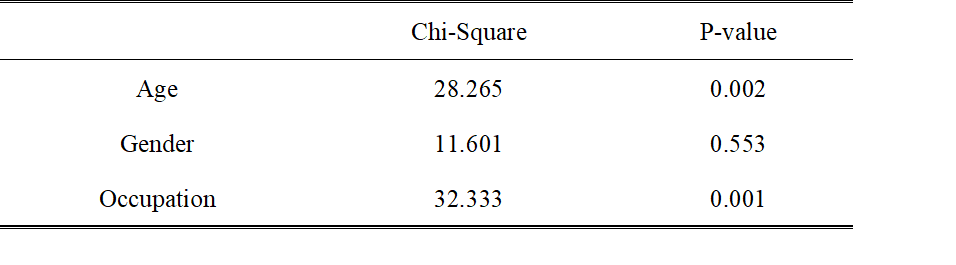}	
		
	\end{table}

\begin{figure}[H]
	\centering
   		
		\includegraphics[width=4in]{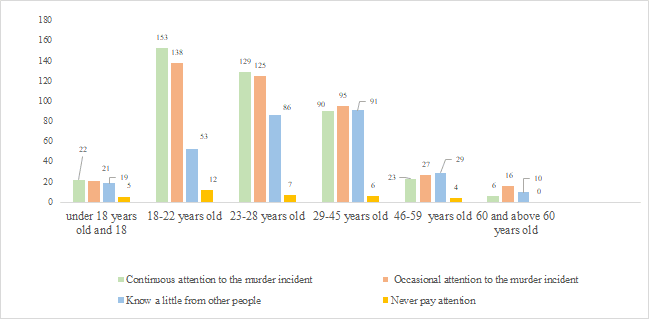}	
\caption{ Age and Incident Attention }
		
		  \label{mm}  
     
	\end{figure}

We can be seen from the figure \ref{mm}, people aged 18-22 have the highest attention; followed by people aged 23-28 have the highest attention. The ride-hailing software is generally used by young people, as well as some middle-aged people. In addition, most of the victims of the incident are young women, hence people aged 18-28 pay more attention to it. This age group is the main user of ride-hailing, including college students, young workers. Therefore these findings are consistent with the facts.

\begin{figure}[H]
	\centering
    
		\includegraphics[width=4in]{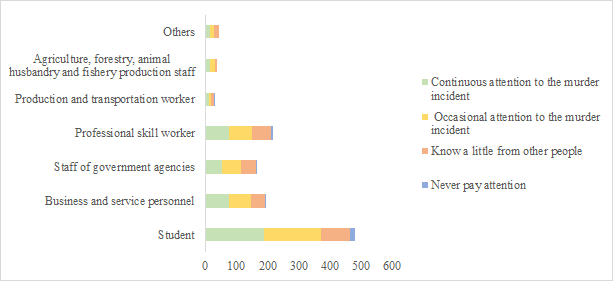}	
		\caption{ Occupation and Incidents Attention }
		
		 \label{po}
	\end{figure}

 We can be seen from the figure \ref{po}, production and transportation workers have the highest attention (40.54\% of their total number), followed by students (37.58\% of their total number). The Incident has direct impact on the use of ride-hailing, and also has a lot of impacts on many drivers, so drivers will pay more attention to the follow-up of the murder incident. Young people is also the main group uses the ride-hailing, therefore students pay attention to the incident. These findings are consistent with the facts.

\subsection{Murder incidents Cause Analysis}

\begin{figure}[H]
	\centering
     
     \label{ Murder Incident Cause }
		\includegraphics[width=4in]{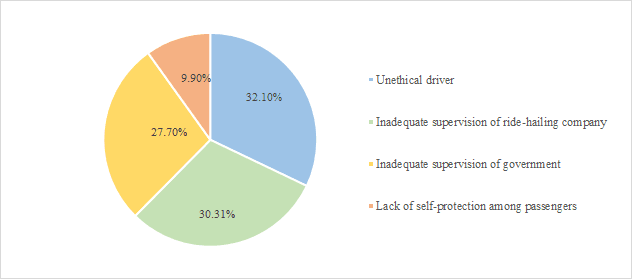}	
		\caption{ Murder Incident Cause }
		
		 \label{ Murder Incident Cause }
	\end{figure}

From the figure\ref{ Murder Incident Cause },32.1\% of people believe that driver's unethical behavior is the main cause of accidents. At the same time, 30.31\% believe that the company's supervision is not in place. 27.7\% of people think that government supervision is inadequate.This shows that ride-hailing companies and the government should strengthen supervision and take necessary measures to reduce such incidents.

\subsection{Impact on the use frequency}
We want to investigate whether there will be an impact on ride-hailing use after the murder  incident, and difference in use frequency under different ages, genders, and occupations. Therefore, we exploit Chi-square to test whether there is a significant relationship between personal characteriastics and use frequency after the extreme incident. Table \ref{bb} shows that the p-value of gender is less than 0.05 at a significant level of 5\%, therefore people are different in gender and the use of ride-hailing varies in frequency.

\begin{table}[H]
	\centering
    
    	\caption{Use Frequency Chi-square Test }

		 \label{bb}
		\includegraphics[width=4in]{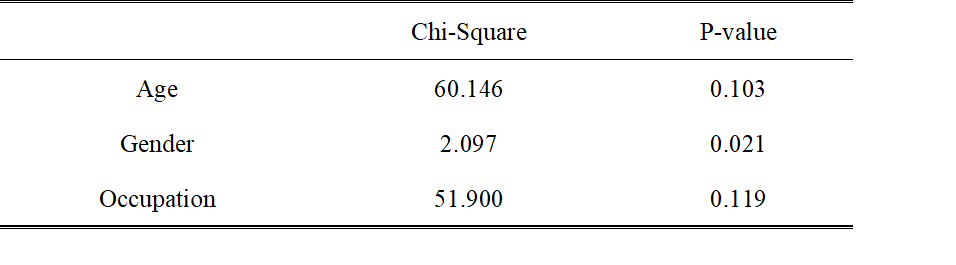} 
	
	\end{table}

\begin{figure}[H]
	\centering

		\includegraphics[width=4in]{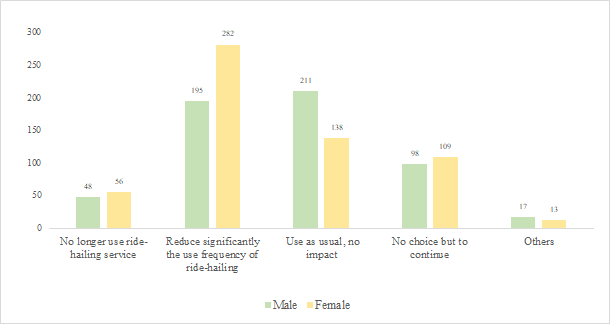} 
		\caption{ Use Frequency after Murder Incidents 
		in Gender }

		 \label{ Use Frequency after Murder Incidents in Gender }
	\end{figure}

The figure \ref{ Use Frequency after Murder Incidents in Gender } presentsd that 282 women(47.47\% of  total number of women) and 195 men (34.2\% of  total number of men) suggested that they would reduce significantly the use frequency of ride-hailing  after the murder incidents; followed by 138 women (accounting for 23.8\% of the total number of women), 211 men (37.08\% of  total number of men) said that they were not affected and used the ride-hailing service as usual; A small number of men and women said they would no longer use ride-hailing.\\

The series of incidents of the ride-hailing passenger killings of  Chinese 21-year-old Air Stewardess killed by Ride-hailing Driver and Chinese 20-Year-Old Woman Raped and Killed by Ride-hailing Driver has aroused widespread concern from all walks of life and people's concerns. Many people have begun to reduce the frequency of using ride-hailing service and take the subway or other means of transportation instead, especially, vulnerable women are even more afraid to use ride-service, such as Didi Chuxing, while men are less affected.

\newpage
\section{  Paired Samples Test for Ride-hailing Safety rectification Impact Analysis}

\begin{figure}[H]
	\centering
    
		\includegraphics[width=5in]{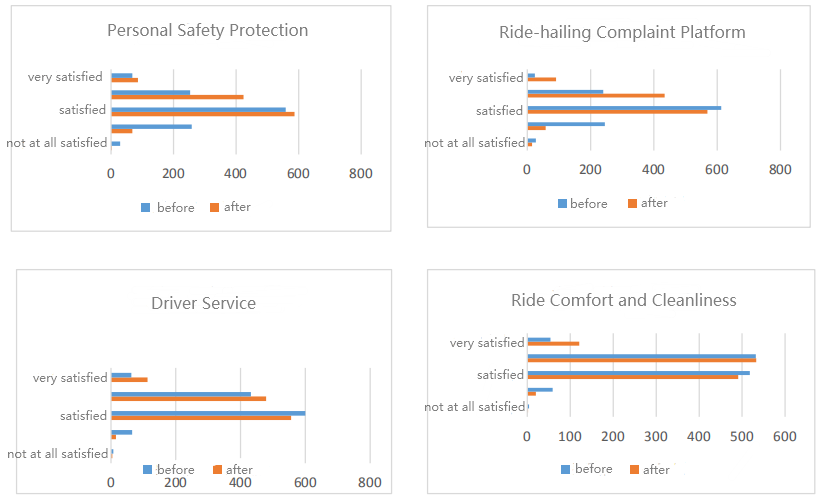}	
		\caption{ Satisfaction change }
		
		 \label{Satisfaction}
	\end{figure}

After a series of murder incidents of ride-hailing drivers, ride-hailing companies took measures to prevent such extreme incidents,including video and audio recording during the trip,improving the standards for selecting drivers,setting up emergency contacts and increasing customer service staff. In order to find whether the customer's satisfaction has been improved after the ride-hailing safety rectification, we asked respondents to give a satisfaction score for each service before and after the rectification(score  from 1 to 5 indicate Not at all Satisfied,Partly Satisfied, Satisfied, More than Satisfied,Very Satisfied). Through a comparative analysis of two samples, we find that rectification significantly affects customers’ satisfaction, and satisfaction score is approximately considered to follow a normal distribution. Therefore, we exploit Paired Sample T-Test to analyze whether these measures have obvious effects.Table \ref{hh }  illustrates that the P-value of each service rectification(except for APP Performance) is less than 0.05 at a significant level of 5\%, which shows that the degree of satisfaction for ride-hailing services has increased. 
\begin{table}[H]
\centering
\caption{ Paired Samples Test for Service Satisfaction}
\label{hh }
\resizebox{\textwidth}{25mm}{
\begin{tabular}{cccccccccc}
		
		\hline\hline
                       
     \multicolumn{2}{c}{\multirow{2}*{}}&
 \multicolumn{8}{c}{95\% Confidence Interval of the Difference}\\
\cline{3-10}

\multicolumn{2}{c}{}&Mean&Std. Deviation&Std. Error Mean&Lower&Upper&t&df&Sig. (2-tailed)\\
\hline
Pair 1 & \tabincell{c}{Personal Safety Protection before -\\after Service  Improvement} &-0.386&	0.746	&0.022	&-0.429	&-0.344	&-17.700&	1167&	0.001\\
Pair 2&	\tabincell{c}{ APP Performance before -\\after  Service  Improvement}&	0.001	&	0.687	&	0.020	&	-0.039	&	0.040	&	0.043	&	1167	&	0.966\\

Pair 3	&\tabincell{c}{	Ride Expenses before-\\after  Service  Improvement	}&	-0.215	&	0.662	&	0.019	&	-0.253	&	-0.177	&	-11.099	&	1167	&	0.000	\\
Pair 4	&\tabincell{c}{	Driver Service before -\\  after Service  Improvement}	&	-0.174	&	0.692	&	0.020	&	-0.214	&	-0.134	&	-8.592	&	1167	&	0.000	\\
Pair 5	&\tabincell{c}{	Ride Comfort and Cleanliness before -\\ after Service  Improvement}	&	-0.093	&	0.702	&	0.021	&	-0.134	&	-0.053	&	-4.544	&	1167	&	0.000	\\
Pair 6	&\tabincell{c}{	Ride-hailing Complaint Platform before -\\ after  Service  Improvement}	&	-0.462	&	0.732	&	0.021	&	-0.504	&	-0.420	&	-21.567	&	1167	&	0.000	\\

	\noalign{\smallskip}\hline\hline
	
	\end{tabular}
	}
	\end{table}
The figure \ref{Satisfaction} shows the change of the customer's satisfaction obviously.Customers feel more satisfied with personal safety protection,complaint platform,driver service and ride comfortable and cleanliness.This change means the measures taken by companies lead a positive influence to customers.
	
\newpage
\section{Factors Influencing Use of Ride-Hailing Services}
 Factors Influencing Use of Ride-Hailing Services
In order to find out the factors that affect people's use of ride-hailing, we exploit Question 19(attitude towards the importance of various performances of ride-hailing software) in the questionnaire to perform Factor Analysis. There are 10 variables in the question 19: App convenience($X_{1}$), Vehicle configuration($X_{2}$), Driver service attitude($X_{3}$), Time spent to call ($X_{4}$), Ride Expenses($X_{5}$), Complaint feedback platform ($X_{6}$), Passenger safety protection ($X_{7}$), Passenger private information protection ($X_{8}$), Driver registration standard ($X_{9}$), Driver identity( $X_{10}$).
\begin{table}[H]
\centering
\caption{Variables }
\label{Variables}
\resizebox{50mm}{22mm}{
\begin{tabular}{c|c }
		
		\hline\noalign{\smallskip}
		Variables& \quad\\
                                     \hline
                            App convenience& $X_{1}$ \\

Vehicle configuration&$X_{2}$\\
Driver service attitude& $X_{3}$\\
Time spent to call &$X_{4}$\\
Ride Expenses&$X_{5}$\\
 Complaint feedback platform& $X_{6}$\\
Passenger safety protection &$X_{7}$\\
Passenger private information protection &$X_{8}$\\
Driver registration standard &$X_{9}$\\
Driver identity &$X_{10}$\\
\hline
		\end{tabular}
	}
	\end{table}

\subsection{KMO and Bartlett’s Test }
Before performing Factor Analysis, data set must be tested by KMO and Bartlett’s Test to examine whether the data is suitable for factor analysis. The Kaiser-Meyer-Olkin (KMO) test measures the applicability of our data to factor analysis. KMO values less than 0.6 indicate the sampling is not adequate that factor analysis should not be taken. The statistics of Bartlett’s Test of Sphericity is derived from the determinant of the correlation coefficient matrix. Small p-value (less than 0.05) of the significance level indicate that factor analysis may be useful with data. Table \ref{ff} shows that the p-values of Bartlett’s Test are less than 0.05 and KMO vaules are more than 0.6 which shows our data is suitable for factor analysis.

  \begin{table}[H]
	\centering
   \caption{KMO and Bartlett’s Test }

     \label{ff}
		\includegraphics[width=4in]{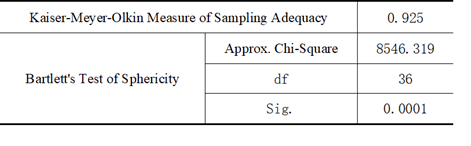}	
		
	\end{table}

 \subsection{Extract Factors}  
\begin{figure}[H]
	\centering
   
		\includegraphics[width=2.5in]{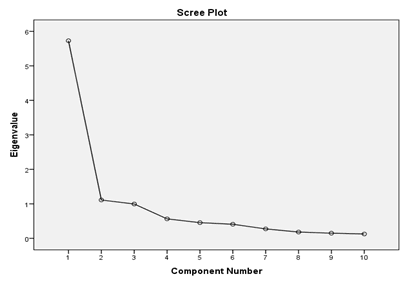}	\caption{ Scree Plot }

     \label{kk}
	\end{figure}

 We can see from table\ref{vv} and figure \ref{kk}  that we can choose 2-4 factors and the first four factor Cumulative percentage of variance  reached 84.589\%( about 85\%), ,variables can be explained well when cumulative percentage of variance reaches 85\% ,therefore we extract four factors.
	
\begin{table}[H]
	\centering
   	\caption{ Total Variance Explained }

     \label{vv}
		\includegraphics[width=2.5in]{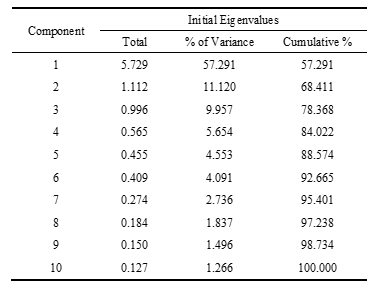}
	
	\end{table}

\subsection{Component Matrix}	
The table \ref{78} below shows the loadings of the ten variables on the four factors extracted. The higher the absolute value of the loading, the more the factor contributes to the variable (we have extracted four variables among the 10 items are divided into 4 variables according to most important items which similar responses in component 1 and simultaneously in component 2 ,3, and 4 ). The gap (empty spaces) on the table represent loadings that are less than 0.5, this makes reading the table easier. We suppressed all loadings less than 0.5 (Table \ref{78}). There are ten variables with high loadings on the first factor and low loadings on  other three factors. In order for each factor to fully explain variables, we rotate the factor.

\begin{table}[H]
\centering

\centering
\caption{Component Matrix}
\label{78}
\resizebox{110mm}{20mm}{

\begin{tabular}{ccccc}

		\hline\hline\noalign{\smallskip}
	&Component1	&	Component2	&	Component3	&	Component4\\
		\noalign{\smallskip}\hline\noalign{\smallskip}
	App convenience	&	0.531	&	0.661	&		&		\\
Vehicle configuration	&	0.645	&	0.518	&		&		\\
Driver service attitude	&	0.758	&		&		&		\\
Time spent to call	&		&		&	0.964	&		\\
Ride Expenses	&	0.736	&		&		&  -0.576		\\
complaint feedback platform	&	0.857	&		&		&		\\
Passenger safety protection	&	0.874	&		&		&		\\
Passenger private information protection	&	0.911	&		&		&		\\
Driver registration standard	&	0.885	&		&		&		\\
Driver identity	&	0.895	&		&		&		\\

		\noalign{\smallskip}\hline\hline
	
	\end{tabular}
	}
	\end{table}

\subsection{ Component Matrix Rotation}
We use the maximum variance method for factor rotation. The first factor explains the complaint feedback platform, user personal security protection, user information security protection, driver registration standards, and driver identity to higher degree, we name it as Safety factor. The second factor has a high degree of interpretation of APP performance and vehicle configuration, we summarized as Setting Factor; the third factor has a high degree of interpretation of driver service attitude, ride expenses, and is summarized as Service Factor; the fourth factor has a high degree of interpretation of Time spent to call, which is summarized as Efficiency Factor. Therefore, we extract four main factors that affect people's use of ride-hailing: Setting Factor, Safety factor, Service factor,  Efficiency Factor.

\begin{table}[H]
\centering
\caption{Rotated Component Matrix}
\label{Component Matrix}
\resizebox{100mm}{22
mm}{
\begin{tabular}{ccccc }
		
		\hline\hline\noalign{\smallskip}
	&Component1	&	Component2	&	Component3	&	Component4\\
	\noalign{\smallskip}\hline\noalign{\smallskip}
App convenience	&		&	0.878	&		&		\\
Vehicle configuration	&		&	0.764	&		&		\\
Driver service attitude	&		&		&	0.641	&		\\
Time spent to call	&		&		&		&	0.999	\\
Ride Expenses	&		&		&	0.829	&		\\
complaint feedback platform	&	0.779	&		&		&		\\
Passenger safety protection	&	0.866	&		&		&		\\
Passenger private information protection	&	0.880	&		&		&		\\
Driver registration standard	&	0.888	&		&		&		\\
Driver identity	&	0.890	&		&		&		\\

		\noalign{\smallskip}\hline\hline
	
	\end{tabular}
	}
	\end{table}

\section{Logistic Regression Analysis on People’s Willingness to Ride}

After the safety rectification, we want to analyze whether people are willing to continue to use ride-hailing service by using Logistic Regression Analysis. We use variables of Question 20(App convenience, Vehicle configuration	Driver service attitude, Time spent to call, 	Fares,	complaint feedback platform,	Passenger safety protection,	Passenger private information protection	Driver registration standard,	Driver identity) and gender, age, occupation as Independent variables, Use Ride-hailing Willingness as dependent variable(YES=1, NO=0)\\

\begin{table}[H]
\centering
 \caption{Logistic Regression Variables }
\label{Variables}
\resizebox{50mm}{22mm}{
\begin{tabular}{c|c }
		
		\hline\noalign{\smallskip}
		Variables& \quad\\
                                     \hline
                                     
     Use Ride-hailing Willingness & $y$  \\                            
                        App Performance & $X_{1}$ \\

Vehicle configuration&$X_{2}$\\
Driver service attitude& $X_{3}$\\
Time spent to call &$X_{4}$\\
Ride Expenses&$X_{5}$\\
 Complaint feedback platform& $X_{6}$\\
Passenger safety protection &$X_{7}$\\
Passenger private information protection &$X_{8}$\\
Driver registration standard &$X_{9}$\\
Driver identity &$X_{10}$\\
Age &$X_{11}$\\
Gender &$X_{12}$\\
Occupation &$X_{13}$\\
\hline
		\end{tabular}
	}
	\end{table}

We can see from Table \ref{gg}  that P-value of Vehicle configuration($x_{2}$),	Ride Expenses($x_{5}$), complaint feedback platform($x_{6}$), Passenger security protection($x_{7}$), Passenger private information protection($x_{8}$), Gender($x_{11}$), Age($x_{12}$) are less than 0.05 at a significant level of 5\%, therefore we remove variables with p-value more than 0.05.\\

\begin{table}[H]
\centering
\caption{ Logistic Regression Result}
\label{gg}
\resizebox{120mm}{23mm}{
\begin{tabular}{ccccccc }
		
		\hline\hline\noalign{\smallskip}

	&	B	&	S.E.	&	Wald	&	df	&	Sig.	&	Exp(B)	\\
\noalign{\smallskip}\hline\noalign{\smallskip}

Gender	&	0.468	&	0.128	&	13.360	&	1	&	0.000	&	1.597	\\
Age	&	-0.247	&	0.070	&	12.618	&	1	&	0.000	&	0.781	\\
Occupation	&	0.058	&	0.048	&	1.466	&	1	&		&	1.060	\\
App convenience	&	0.144	&	0.132	&	1.192	&	1	&		&	1.155	\\
Vehicle configuration	&	-0.300	&	0.128	&	5.514	&	1	&	0.019	&	0.741	\\
Driver service attitude	&	0.198	&	0.150	&	1.742	&	1	&		&	1.218	\\
Time spent to call	&	0.112	&	0.090	&	1.545	&	1	&		&	1.118	\\
Ride Expenses	&	0.348	&	0.118	&	8.746	&	1	&	0.003	&	1.416	\\
complaint feedback platform	&	-0.280	&	0.122	&	5.308	&	1	&	0.021	&	0.756	\\
Passenger safety protection	&	-0.595	&	0.139	&	18.335	&	1	&	0.000	&	0.551	\\
Passenger private information protection	&	0.318	&	0.150	&	4.513	&	1	&	0.034	&	1.375	\\
Driver registration standard	&	0.116	&	0.157	&	0.548	&	1	&		&	1.123	\\
Driver identity	&	-0.101	&	0.148	&	0.467	&	1	&		&	0.904	\\
Constant	&	-0.516	&	0.597	&	0.745	&	1	&		&	0.597	\\

	\noalign{\smallskip}\hline\hline
	
	\end{tabular}
	}
	\end{table}

From Table \ref{LL},we  we can obtain logistic regression equation:
 \begin{equation}
  \begin{split}
Logit=\ln{\frac{p}{1-p}}=-0.115X_{2}+0.386X_{5}-0.175X_{6}-0.616X_{7}+0.347X_{8}+0.458X_{11}(2)+\\1.141X_{12}(1)+1.248X_{12}(2)+0.87X_{12}(3)+1.172X_{12}(4)-0.192X_{12}(5)
    \end{split}
\end{equation}

\begin{table}[H]
\centering
\caption{Final Logistic Regression Result}
\label{LL}
\resizebox{\textwidth}{27mm}{
\begin{tabular}{ccccccccc}
		
		\hline\hline\noalign{\smallskip}

\multicolumn{7}{c}{}	&	\multicolumn{2}{c}{95\% C.I.for EXP(B)}			\\
\cline{8-9}
	&	B	&	S.E.	&	Wald	&	df	&	Sig.	&	Exp(B)	&	Lower	&	Upper	\\
	\hline
Gender(1)	&	-0.458	&	0.127	&	12.965	&	1	&	0.000	&	0.633	&	0.493	&	0.812	\\
60 and above 60  years old	&		&		&	28.977	&	5	&	0.000	&		&		&		\\
under 18 years old and 18	&	1.141	&	0.529	&	4.649	&	1	&	0.031	&	3.130	&	1.109	&	8.830	\\
18-22 years old	&	1.248	&	0.475	&	6.892	&	1	&	0.009	&	3.482	&	1.372	&	8.837	\\
23-28 years old	&	0.870	&	0.478	&	3.312	&	1	&	0.069	&	2.386	&	0.935	&	6.088	\\
29-45 years old	&	1.172	&	0.479	&	5.989	&	1	&	0.014	&	3.229	&	1.263	&	8.257	\\
46-59  years old	&	-0.192	&	0.553	&	0.121	&	1	&	0.728	&	0.825	&	0.279	&	2.438	\\
Vehicle configuration	&	-0.115	&	0.113	&	1.045	&	1	&	0.307	&	0.891	&	0.715	&	1.111	\\
Fares	&	0.386	&	0.114	&	11.499	&	1	&	0.001	&	1.471	&	1.177	&	1.838	\\
complaint feedback platform	&	-0.175	&	0.120	&	2.108	&	1	&	0.147	&	0.840	&	0.663	&	1.063	\\
Passenger safety protection	&	-0.616	&	0.132	&	21.761	&	1	&	0.000	&	0.540	&	0.417	&	0.700	\\
Passenger private information protection	&	0.347	&	0.133	&	6.751	&	1	&	0.009	&	1.414	&	1.089	&	1.837	\\
Constant	&	-0.691	&	0.622	&	1.236	&	1	&	0.266	&	0.501	&		&		\\

	\noalign{\smallskip}\hline\hline
	
	\end{tabular}
	}
	\end{table}
We can see from table \ref{pp}, taking men as the standard 1, women are 0.633 times more than men to use ride-hailing service even if ride-hailing companies have taken measures to prevent such extreme incidents, this finding is consistent with previous finding that the murder incidents of ride-hailing drivers have more effect on the use of ride-hailing for female. What is more, considering age group 60 and above 60 years old as the standard 1, people aged under 45 are more willing to take a ride, but people aged 46-59 are less willing than 60 and above 60 years old group. P-values are all less than 0.05 that shows our standard is valuable.	

\begin{table}[H]
\centering
\caption{ Logistic Regression Analysis }
\label{pp}
\scriptsize
\begin{tabular}{ccc}
		
		\hline\hline\noalign{\smallskip}

\quad		&	OR(95\% CI)				&	P-value	\\
\cline{2-3}
Gender&					&		\\
Male	&	1				&		\\
Female	&	0.633 (0.493 -0.812) 	&	$\le 0.05$	\\
Age	&					&		\\
\quad\quad60 and above 60  years old	&	1				&		\\
\quad\quad under 18 years old and 18	&	3.130(	1.109-	8.830)	&		\\
18-22 years old	&	3.482(1.372-8.837)	&		\\
23-28 years old	&	2.386(0.935-6.088)	&		\\
29-45 years old	&	3.229(1.263-8.257)	&		\\
46-59  years old	&	0.825(0.279-2.438)	&	$\le 0.05$	\\

	\noalign{\smallskip}\hline\hline
	
	\end{tabular}

	\end{table}
\subsection{Omnibus Tests of Model Coefficient and Hosmer and Lemeshow Test }

We perform Omnibus Tests of Model Coefficient, p-value is less than 0.05 which means that OR value of at least one variable is statistically significant,  the model is great. And the p-value of Hosmer and Lemeshow Test is also less than 0.05 that indicates the information of data has been fully extracted, and the model fits well, result is shown in table  \ref{fd}.

\begin{table}[H]
	\centering
   \caption{Omnibus Tests of Model Coefficient and Hosmer and Lemeshow Test }

     \label{fd}
		\includegraphics[width=4in]{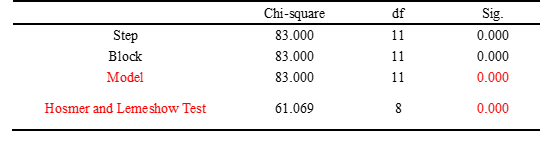}	
	\end{table}

\section{Ride-hailing Market Analysis of Four Quadrant Model }	
\subsection{  Evaluation Indicators Selection }
In order to systematically understand the current users' evaluation of the service and safety guarantee of ride-hailing, and further propose the suggestions from the user group perspective, our study focuses on the construction of “Ride-hailing Evaluation Index System of Security and Service ". The ride-hailing service mainly consists of three parts: online service, offline service, and after-sales service (user complaint, user satisfaction survey, and user feedback). Its service have the following aspects:
\begin{itemize}
\item It is difficult to quantify the service content: the ride-hailing service runs through the entire process of passenger appointments, trip, and payment of fees. It is difficult for regulators and ride-hailing companies to supervise the actual process.
\item  Variability of service evaluation: passengers are an important part of the evaluation system of ride-hailing servic. However, the subjectivity of passengers will have a great impact on the authenticity of the evaluation of ride-hailing service.
\item  Differences in service quality: ride-hailing drivers are the core of the entire offline service. However, the quality and content of services provided by different drivers and different models are also quite different.
\end{itemize}

Based on the consideration of the above three aspects, the index system selects 4 objective aspects : service reliability, service security, service comfort, and service flexibility. Considering the current public opinion for the murder incident, the research decided to highlight several indicators concerning the personal security of users. Then we use ten variables to build Ride-hailing Evaluation System of Security and Service.

\begin{table}[H]
	\centering
     \caption{ Ride-hailing Evaluation System  Variables}
     \label{ Ride-hailing Evaluation System of Security and Service Variables }
		\includegraphics[width=4in]{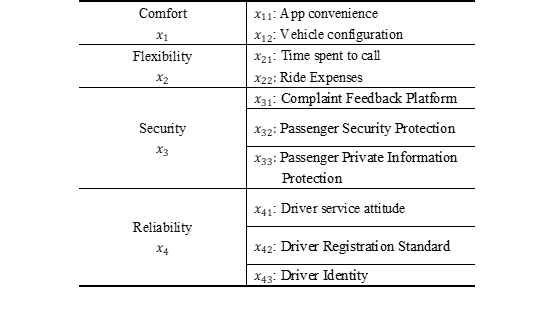}	
	\end{table}

\subsection{Importance Analysis for Ride-hailing Service}

The questionnaire lists ten indicators, allowing users to score the importance of these aspects for ride-hailing company. The scores from 1 to 5 indicate “Not at all Important,” “Partly Important,” “Important,” “More than Important,” “Very Important”. 
\begin{table}[H]
	\centering

		\includegraphics[width=4in]{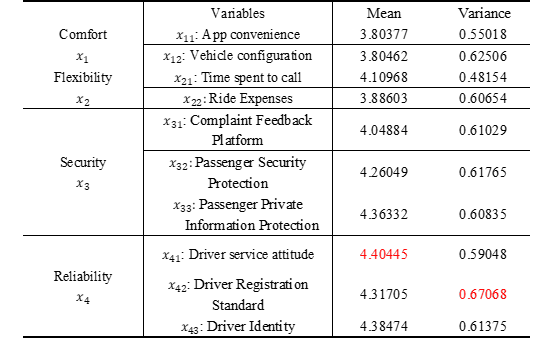}	
		\caption{ Importance Score}
		
		 \label{ Importance Score }
	\end{table}

It can be seen in the figure \ref{ Importance Score } that the means of the ten scores are all more than 3, which indicates that users generally have high requirements for various aspects of ride-hailing service. Specifically,
\begin{itemize}
\item The importance scores of APP performance, vehicle configuration, and ride expenses are all lower than 4, indicating that users have no excessive requirements for comfort and flexibility;

\item The mean of the importance scores of Security exceeds 4.2, a which may be affected by many recent "murder incidents" reports, making people very worry about personal security.

\item According to the variance, the biggest variance is the indicator of "Driver Registration Standard", which shows that people perceptions for it vary widely, or perhaps "driver registration standard" is not widely popularized. It can be seen that the "driver service attitude" has a high mean score and a small variance because people have common perceptions for it.
\end{itemize}

\subsection{Satisfaction analysis for Ride-hailing Service}
The questionnaire lists ten indicators in the questionnaire to score the satisfaction of these aspects for ride-hailing company. The scores from 1 to 5 indicate Not at all Satisfied, Partly Satisfied, Satisfied, More than Satisfied, Very Satisfied. 
\begin{table}[H]
	\centering
     
   \caption{ Satisfaction Score}
		
		  \label{ Satisfaction Score }
		\includegraphics[width=4in]{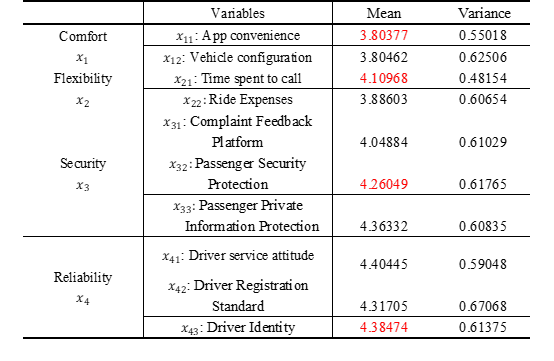}
		
	\end{table}
	
The questionnaire lists ten indicators in the questionnaire to score the satisfaction of these aspects for ride-hailing company. The scores from 1 to 5 indicate Not at all Satisfied, Partly Satisfied, Satisfied, More than Satisfied, Very Satisfied.

The mean of the ten variables are all more than 3, indicating that users are satisfied with the construction of the current ride-hailing platform. The demarcation of the score is also obvious: taking the 3.3 as the boundary, the comfort and flexibility scores are above this, and the safety and reliability scores are below this. Specifically:
\begin{itemize}

\item Both "APP Performance" and " Time Spent to Call" with high satisfaction scores in Comfort and Flexibility  are more than 3.5, indicating that ride-companies have great infrastructure and sufficient car supply.
\item People's high evaluation of "Passenger Security Protection " has further affirmed the technology of ride-hailing company.

\end{itemize}

\subsection{ Importance-Satisfaction Matrix Construction}

In order to give further suggestions for the development of the ride-hailing company, combining importance and satisfaction, our study  use the importance-satisfaction matrix to summarize these ten variables.

\subsubsection{Theoretical background}

The concept of importance-satisfaction matrix comes from a commonly used customer satisfaction research model, the four quadrant model. 
\begin{itemize}
\item When the variable is distributed in the area A (Dominant Area), it shows that both the importance and satisfaction evaluation are high.
\item When the variable is distributed in the area B (Repair Area), it means that these factors are important to customers, but the customer satisfaction evaluation is low, and it is necessary to focus on repair and improvement.

\item When the variable is distributed in the area C (Opportunity Area), it presents that this part of the factor is not the most important for customers, and the satisfaction evaluation is also low, so it is not the most urgent problem to be solved now;

\item When the variable is distributed in the area D (Maintenance Area), it indicates that the satisfaction evaluation is high, but it is not the most important factor for customers.

\end{itemize}

\begin{figure}[H]
	\centering
     
		\includegraphics[width=4in]{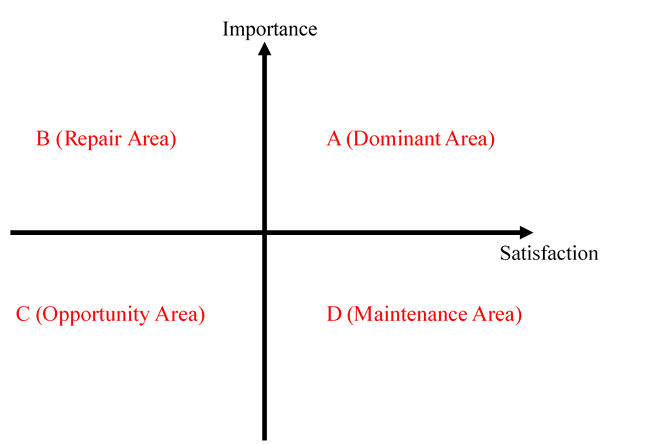}	
		
		\caption{ Importance-Satisfaction Matrix Constructio}
     \label{ Importance-Satisfaction Matrix Construction }
	\end{figure}

\subsubsection{Empirical analysis}
We separate the score data of male and female users, calculate the mean and variance of the importance and satisfaction scores respectively, and obtain the result as shown in the following table \ref{op}.

\begin{table}[H]
	\centering
     	
		\caption{ Overall Mean}
     \label{op}
		\includegraphics[width=4in]{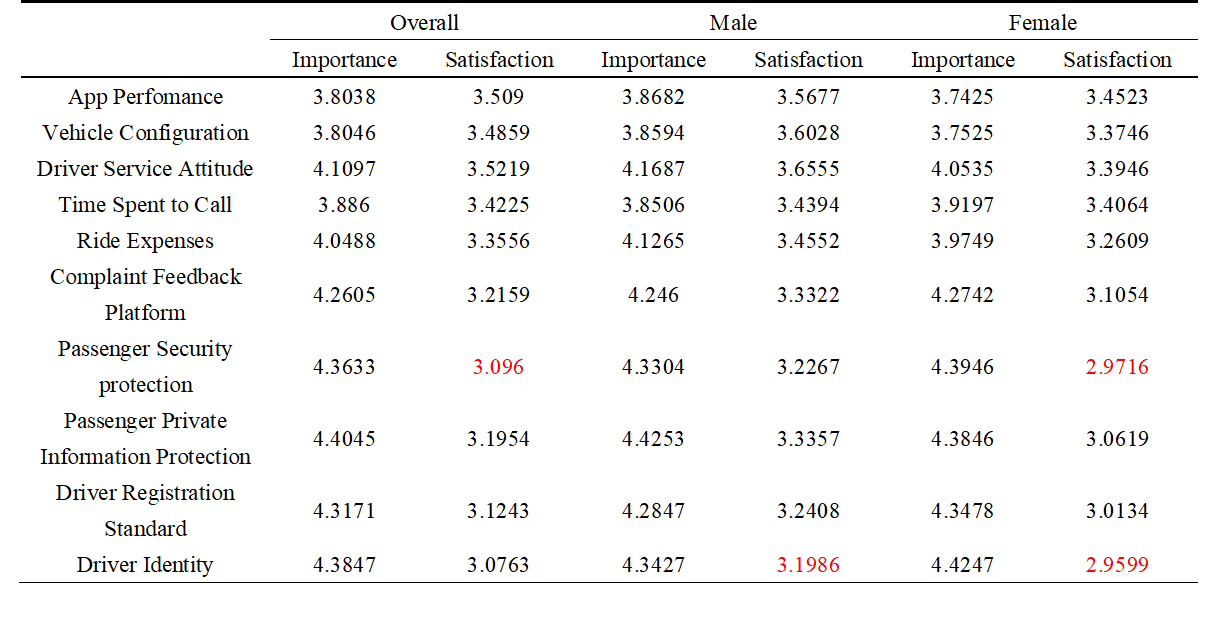}	
	
	\end{table}

We use the mean of importance score and satisfaction as the origin of the x and y axis, the importance score as y axis, satisfaction score as x axis. Table \ref{op} shows that ten variables of security service of ride-hailing are mainly distributed in the Repair Area and the Maintenance Area in general.

\begin{itemize}
\item The variables in the Maintenance Area are: Ride Expenses, Time Spent to call, vehicle Configuration, APP Performance. These variables are related to service efficiency. This shows that ride-hailing market is now generally mature, and the supply and demand in the market are balanced.

\item The variables in the Dominant Area include “ Driver Service Attitude”. Satisfaction of this variable is the highest among the ten variables, and its importance is in the middle level, indicating that most drivers have positive service attitude.
\item The five variables in the repair area are related to security which shows that most users are very concerned with safety issues.

\end{itemize}

\begin{figure}[H]
	\centering
    	
		\includegraphics[width=5in]{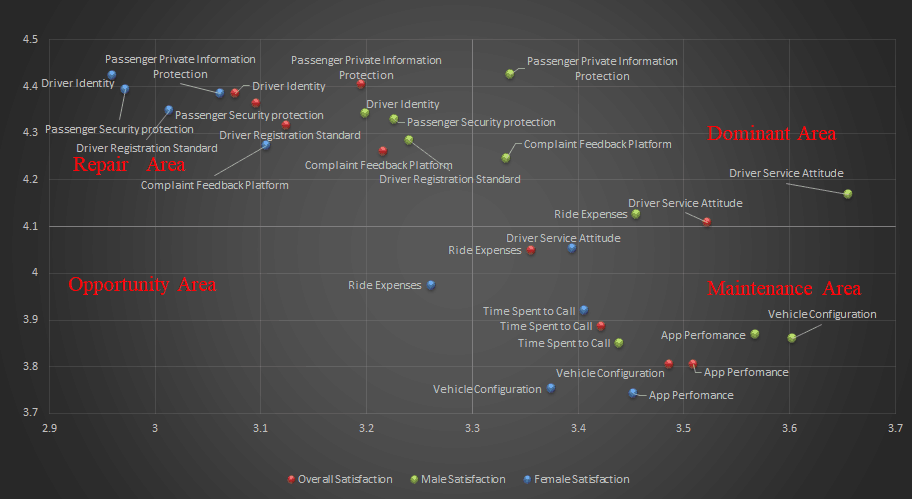}
		\caption{ Four Quadrant Plot}

     \label{nj}
	\end{figure}

Figure \ref{nj} shows that blue dots represent female users, and green dots represent male users. Generally speaking, the scores of female users are on the left, and the scores of male users are on the right, that is, the satisfaction of female users on many service is lower than that of male users.

This makes sense because in the most recent cases, the victims were all women, so women were far more satisfied and trusted with ride-hailing vehicles than men.This also provides the direction for our next improvement, which is to be more caring for the feelings of female users and repair the trust.

Specifically, female users were the least satisfied with driver identity verification, user personal safety protection and driver registration standards.Men were also the least satisfied in all three categories, with an average satisfaction score about 0.2 points higher than women.Both male and female users expressed concern about the qualification of drivers and the personal safety protection of users.This may be the main direction of future improvement.

\subsection{ Construction of evaluation index system for safety service level of online ride-hailing vehicles}

\subsubsection{Determine the Weight}

As there are many qualitative contents of online ride-hailing services in the above mentioned, the relations among the contents overlap and influence each other, and the evaluation indexes that belong to a certain sub-index should not only reflect the information of the sub-index, therefore, the research decides to further establish a complete model in view of this situation.

Due to the different influences of each indicator on the final result in the ride-hailing service, different weights should be assigned according to the reality.Due to the particularity of ride-hailing service, entropy weight method is used to weight the importance of each evaluation index of established ride-hailing service level.

The calculation process of entropy weight method is as follows:

\begin{itemize}
\item Construct judgment matrix.
$\mathrm{R}=\left(X_{i j}\right)_{m n}(i=1,2, \ldots, \mathrm{m}, \mathrm{j}=1,2, \ldots, \mathrm{n})$
 Among them,m refers to the number of respondents and n refers to the number of indicators. The importance score of each respondent's ten indicators in turn for each behavior.
 
\item Normalized and get the judgment matrix B.
It is known that there are four methods to determine the degree of subordination. Here, combined with the characteristics of ride-hailing service, the larger the better mode is selected, and the formula is as follows:
$b_{i j}=\frac{x_{i j}-\min x_{i j}}{\max x_{i j}-\min x_{i j}}$

\item Calculate the entropy formula.
$H_{j}=\frac{\left(-\sum_{i=1}^{m} f_{i j} \ln f_{i j}\right)}{\ln m}$ 
   and   $f_{i j}=\frac{1+b_{i j}}{\sum_{i=1}^{m}\left(1+b_{i j}\right)}$

\item Calculate the weight formula.
$\omega_{j}=\frac{\left(1-H_{j}\right)}{n-\sum_{j=1}^{n} H_{j}}$

\item Through the calculation of the development capacity coefficient weight.

\begin{table}[H]
\centering
\caption{ Index entropy weight distribution table }
\label{weight}
\resizebox{80mm}{25mm}{
\begin{tabular}{cc}
		
		\hline\hline\noalign{\smallskip}

\     Index		&	Entropy Weight	\\
\noalign{\smallskip}\hline\noalign{\smallskip}

APP Performance&  0.082259618 \\
Vehicle Hardware Configuration&  0.09925608 \\
Driver Service Attitude&  0.102440109 \\
Time and Success Rate of Car Booking&  0.095682743 \\
Bus Fee&  0.075473444 \\
Evaluate the Complaint Feedback System&  0.102975938 \\
Personal Safety Protection for Users&  0.1 15305939 \\
User Information Security Protection&  0.105922674 \\
Driver Registration Criteria&  0.110382938 \\
Driver Identification Review&  0.110300519 \\

	\noalign{\smallskip}\hline\hline
	
	\end{tabular}
	}
	\end{table}
	
\end{itemize}

From the obtained weights, it can be seen that the weights of comfort and flexibility are relatively low, and the weights of safety and reliability are relatively high. However, the weight difference of each evaluation index is not large, which is basically around 1, which is in line with the convention.When improving the construction of the platform, the indicators with a large weight and those related to safety should be given priority.

\subsubsection{Results Analysis}

The user satisfaction matrix is denoted as:$M_{Y}$,this is an m by n matrix.The index entropy weight $W_{j}$ of n evaluation indexes can form n×1 matrix W.The m×1 matrix R of the sample entropy value of all samples can be obtained by multiplying the two matrices, which can be calculated as follows:
$R=M_{Y} \times W$
The adjustment makes the overall score 1 when the user scores all the ten indicators 1.When the user scores 5 for all the 10 indicators, the comprehensive score is 5, and then the obtained weight is used to calculate the comprehensive satisfaction of each user as follows:

\begin{figure}[htp]
    \centering
    \includegraphics[width=12cm]{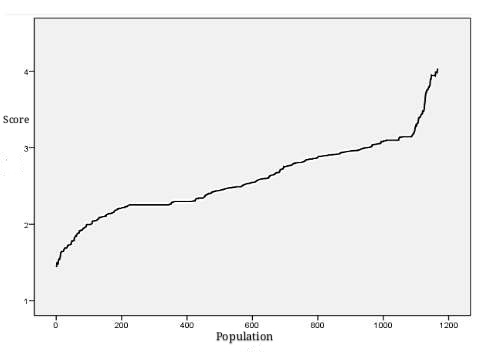}
    \caption{User comprehensive score distribution map}
    \label{fig:score}
\end{figure}

It can be seen that the scores are concentrated in 2-3 points, and there are only about 100 users with scores above 3 points, which is generally low, indicating that the overall satisfaction of users to online ride-hailing vehicles is low, and the industry as a whole urgently needs improvement.

\subsection{ Chapter Summary}

Most ride-hailing users have a high evaluation on the comfort and flexibility index, but they think it is of low importance. Safety and reliability were rated low, but their importance was considered high. To the credit of the net about car platform mature technology and high efficiency, which brought a lot of convenience for the user's life, but there is also security vulnerabilities, and has caused a number of personal injury, through the batch report is further brought panic and mental harm to the user, subtly changed the user's attitude to network about car and evaluation, the extreme distrust platform security, expect to change as soon as possible.Such "psychological harm" is more obvious among female users. Ride-hailing platforms need to fix not just security loopholes, but also a widening rift in trust between users, platforms and drivers.

\section{Conclusion}
Through Chi-square Test and Cross Analysis, we found that murder incidents have adverse impact on women, but not much on men. We quantified women's willingness to use ride-hailing service was 0.633 times that of men by using Logistic Regression . This is consistent with the fact that most of the victims of murders are women, and women are at a disadvantaged, making most women unwilling to continue to use ride-hailing service.Then, we extract four main factors that affect people's use of ride-hailing:Setting Factor, Safety Factor, Service Factor, Efficiency Factor.\\
\\
Most ride-hailing users’ evaluation of Comfort and Flexibility is higher, but people consider it less important. On the other hand, people are less satisfied with safety and reliability of ride-hailing service, they think it is very important. The high efficiency of the ride-hailing platform has brought convenience to users' lives. However, personal danger has always existed and caused murder incidents of ride-hailing drivers. Social media report for murder incidents  brought panic to the public, and influenced user's attitude and evaluation of the ride-hailing service subjectively. Such "mental harm" is more obvious to female users. The ride-hailing company not only should fix security vulnerabilities, but also should build trust among users, platforms and drivers.

\end{document}